\documentclass[paper]{ieice}
\usepackage{amsmath}
\usepackage{amssymb} 
\usepackage{graphicx}
\usepackage{latexsym}
\usepackage[varg]{txfonts}
\usepackage{multirow}

\DeclareMathOperator*{\argmaxB}{argmax}
\def\ClassFile{\texttt{ieice.cls}}

\newcommand{\AmSLaTeX}{%
 $\mathcal A$\lower.4ex\hbox{$\!\mathcal M\!$}$\mathcal S$-\LaTeX}
\def\BibTeX{{\rmfamily B\kern-.05em
 \textsc{i\kern-.025em b}\kern-.08em
  T\kern-.1667em\lower.7ex\hbox{E}\kern-.125emX}}
\makeatletter
\def\tmpcite#1{\@ifundefined{b@#1}{\textbf{?}}{\csname b@#1\endcsname}}%
\makeatother

\field{E}
\vol{93}
\no{1}
\SpecialSection{Recent Advances in Machine Learning for Spoken Language Processing}
\title[How to Use the Class File (\ClassFile)]
      {Noise Robust Speech Recognition Using Multi-Channel Based Channel Selection And Channel Weighting}
\authorlist{%
 \authorentry[s147002@stn.nagaokaut.ac.jp]{Zhaofeng ZHANG}{n}{labelA}
 \authorentry[XIAOXIONG@ntu.edu.sg]{Xiong XIAO}{n}{labelB}
 \authorentry[wang@vos.nagaokaut.ac.jp]{Longbiao WANG}{m}{labelA}
 \authorentry[ASESChng@ntu.edu.sg]{EngSiong CHNG}{m}{labelB}
 \authorentry[hli@i2r.a-star.edu.sg]{Haizhou LI}{m}{labelB}
}
\affiliate[labelA]{The author is with the Faculty of Nagaoka University of Technology}
\affiliate[labelB]{The author is with the Faculty of Nanyang Technological University}
\received{2016}{2}{4}
\revised{2016}{2}{4}
\begin{document}
\maketitle

\begin{summary}
In this paper, we study several microphone channel selection and weighting methods for robust 
automatic speech recognition (ASR) in noisy conditions. For channel selection, we investigate 
two methods based on the maximum likelihood (ML) criterion and minimum autoencoder reconstruction 
criterion, respectively. For channel weighting, we produce enhanced log Mel filterbank coefficients 
as a weighted sum of the coefficients of all channels. The weights of the channels are estimated 
by using the ML criterion with constraints. We evaluate the proposed methods on the CHiME-3 noisy ASR task. 
Experiments show that channel weighting significantly outperforms channel selection due to its 
higher flexibility. Furthermore, on real test data in which different channels have different 
gains of the target signal, the channel weighting method performs equally well or better than 
the MVDR beamforming, despite the fact that the channel weighting does not make use of the phase 
delay information which is normally used in beamforming. 
\end{summary}

\section{Introduction}
\label{intro}
The performance of the state-of-the-art automatic speech recognition (ASR) 
systems degrades significantly in far field scenarios due to the existence 
of noise and reverberation \cite{c1}. In many applications, a close-talking recording 
device is not possible, hence many methods have been investigated to reduce 
the effect of noise and reverberation \cite{c10,c11,c27,c12}. 
In this paper, we focus on using multiple microphones for robust ASR.


When microphone array signals are avalabile, a popular way to enhance the source signal is to apply 
beamforming that performs spatial filtering to reduce noise and reverberation \cite{c1,c3,c16,c28}.
 The minimum variance distortionless response (MVDR) beamformer is one of the popular beamforming algorithm
 \cite{c1,c4,c26}.
 The MVDR beamformer fixes the gain of the desired direction to unity while minimizing the 
 variance of the output signal, hence it does not cause distortion to the target signal if the direction of 
 target signal is estimated correctly. Beamforming usually performs well when the signal qualities 
 in the microphone channels are similar. However, in many real applications, the signal qualities 
 of channels can be very different, and beamforming may not be able to improve ASR performance \cite{c5}.

An alternative way to beamforming is to select the channel with the best signal quality for speech recognition.
 Several channel selection methods have been proposed \cite{c6,c7,c18,c19}. For example, the signal-to-noise ratio 
 (SNR) based method selects the channel with the highest SNR \cite{c6}. 
 Another position based method \cite{c7} chooses 
 the microphone channel closest to the speaker which is believed to have the least distortion. The closest 
 channel can be identified using the time delay of arrival information of the channels. Although the SNR 
 and position based method are simple and computationally efficient, the SNR estimation may not be accurate 
 in real applications and the nearest microphone cannot always guarantee the best ASR performance. 
 
In this paper, we present two channel selection methods by using prior information of clean speech. 
In the first method, we choose the channel whose features have the maximum likelihood when 
evaluated on a Gaussian mixture model (GMM) trained on clean features. In the second method, 
we choose the channel with minimum reconstruction error with the clean-trained autoencoder \cite{c8,c29}.
This is motivated by the hypothesis that the channel with the smallest reconstruction
 error has the best fit to the clean-trained autoencoder. 
 
A limitation of channel selection is that only one channel's information is used for ASR, 
hence the information in the array signals is not fully exploited. To address this limitation, we consider 
using all channels' information instead of selecting one channel.
 Besides traditional beamforming, there has been several studies on 
 fusing the information of multiple channels. For example, in \cite{c15}, 
 which motivated by the multi-stream/multi-band ASR strategy \cite{c18}, 
 the authors divide the signal spectrum into subbands and perform beamforming 
 in each subband. A set of HMMs are trained for each subband, and the information 
 is fused at the likelihood score domain. In another study \cite{c17}, the signals 
 received by a distributed microphone array are fused at the time domain through a 
 linear weighted sum to recognize song types. 

In this paper, we propose to fuse the channels in the log Mel filterbank domain 
for ASR. Specifically, the filterbanks of channels are weighted and summed to produce 
a single set of enhanced filterbanks that are used for acoustic modeling. Ideally, if 
a channel's signal quality is bad, it will have a small weight in the fusion and vice versa.
 We have called the proposed method channel weighting. Channel selection can be seen as a special case of channel 
 weighting where the selected channel has a weight of 1 and the rest channels have
 weight 0. A ML-based objective function is proposed for estimating the channel weights. Similar
 to channel selection method, a GMM is trained with clean speech. Then we estimate
 the weights of channels which gives ML on this model. To improve the ML-based weight 
 estimation, two constraints are imposed on the weight estimation problem. In the first
 constraint, the sum of the weights is normalized to 1 after being estimated. In the second 
 constraint, the log determinant of the covariance matrix of the output filterbanks is also considered
 in the ML objective function.

The rest of this paper is organized as follows. Section 2 describes the mechanism of channel
 selection methods. Section 3 defines the formulation of our channel weighting problem and the
 two constraints. Section 4 evaluates these methods on an ASR system based on CHiME-3 \cite{c9} speech data set.
 Conclusions of this work and future plans are summarized in section 5.

\section{Channel Selection Method}
\label{usage}
In this study, the channel selection plays a important role as a part of ASR system front-end. The channel with 
the best speech quality is chosen from microphone channels. The performance of channel selection is 
measured by the word error rate (WER) of ASR's system. Note that the presented channel selection methods 
are applied on feature domain, i.e. on the log Mel filterbank features of 40 dimensions.

\subsection{Maximum likelihood based channel selection method}
This channel selection method is based on the idea that the channel 
of better speech quality should have more similar feature distribution
 to that of close-talk clean speech. Given a universal GMM trained from close-talk speech, the log-likelihood $L_c$ of each channel can be obtained by the following equation:
\begin{equation}
 L_c = \frac{1}{T}\sum^T_{t=1}\log p(\textbf{x}_c(t)|\Lambda)
\label{eq:e1}
\end{equation}
where $\textbf{x}_c(t)$ is the feature vector of the $t^{th}$ frame extracted from the
 $c^{th}$ channel�� speech signal. $\Lambda = \{\omega_m, \mu_m, \Sigma_m \}$ is the parameters of the
 universal GMM, where $\omega_m$, $\mu_m$, $\Sigma_m$ are the prior weight, mean vector, and covariance matrix of the $m^{th}$ Gaussian of the GMM which has $M$ Gaussians. 
 $T$ is the number of frames of current utterance. Channel $c$ with the largest $L_c$ will be selected for speech recogntion. 
 \begin{equation}
 c_{ML}^*=\arg\max_{c}L_c, c\in[1,...,C]
\label{eq:e1a}
\end{equation}
where $C$ is the total number of microphone channels for selection. 
 In the GMM training and channel selection, for each channel, feature vectors $\textbf{x}_c(t)$ are processed by mean normalization (CMN) 
 to compensate for the channel distortion. 
 Variance normalization (CVN) is also applied to make the features of all channels
 having the same variance, such that the likelihood function in (\ref{eq:e1}) is more comparable.
 
\subsection{Autoencoder based channel selection method}
\begin{figure}[tb]
\begin{center}
\includegraphics[scale=0.5]{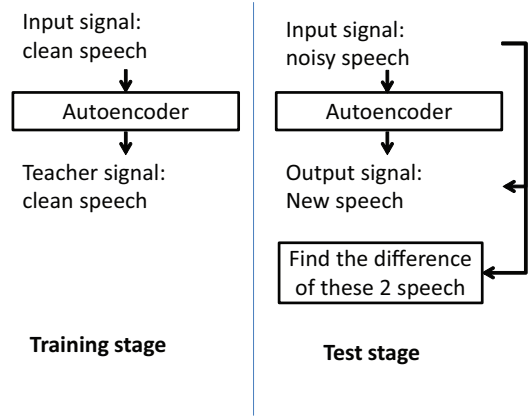}
\end{center}
\caption{The flowchart of autoencoder based channel selection method}
\label{fig:1}
\end{figure}

The motivation of this approach is that a clean speech trained autoencoder is able to reconstruct
 clean speech with small reconstruction error as autoencoder models
 the characteristics of the clean speech \cite{c8,c13}. If noisy speech is
 fed into the autoencoder, the output speech will most likely not be close to the input as
 the autoencoder never sees such data. Hence, it is possible to use the reconstruction error
 produced by the clean trained autoencoder to measure how close
 the input speech is to the clean speech. This approach is illustrated in Fig.~\ref{fig:1}. 

The autoencoder is trained from the same close-talk speech that are used to train the universal GMM
 in the ML based channel selection method. Both the input and teacher signals of the autoencoder are
 the close-talk speech. The best channel selected for ASR can be identified by
 \begin{equation}
 c_{AE}^*=\arg\min_{c}||{\mathcal{X}}_c-f_{AE}(\mathcal{{X}}_c)||^2 , c\in[1,...,C]
\label{eq:e2a}
\end{equation}
where $\mathcal{{X}}_c=[\textbf{x}_c(1)^T,...\textbf{x}_c(T)^T]^T$ is the feature matrix of channel $c$ by concatenating the feature vectors. $f_{AE}()$ represents the nonlinear transformation performed by the autoencoder on the input features, which are CMN processed. 
 
 In this study, the input context size of the autoencoder is 9 frames, hence the total number of inputs
 is 360 dimensions. The autoencoder outputs predicted 40-dimension log Mel filterbanks. There are 3 hidden
 layers each with 1024 hidden nodes.

\section{Channel Weighting Method}
Channel selection only uses the selected channel for ASR and discards the information in the other channels.
 This will limit the performance of the ASR. To address this limitation, 
we study channel weighting in this section, where the features of all 
 channels are weighted and summed to produce the enhanced features. 

The channel weights are estimated by maximizing the log-likelihood of enhanced features on the universal GMM trained from close-talk speech: 

\begin{equation}
\hat{\bf w} = \arg\max_{\textbf{w}} \sum^T_{t=1} \log p({\bf X}_{t}{\bf w}|\Lambda)
\end{equation}
where ${\bf w}$ represents the  $C \times 1$ weight vector for combining the features of the $C$ channels. ${\bf X}_t=[\textbf{x}_1(t)^T,...,\textbf{x}_C(t)^T]^T$ is the  $D \times C$ feature matrix that contains all the feature vectors of $C$ channels at time frame $t$, and  $D$
 is the dimension of the feature vector. To simplify this problem, all dimension of feature shares one 
 weight. ${\bf X}_{t}{\bf w}$ is the linear weighted sum of the features of $C$ channels and used for ASR.  Here the feature vectors are normalized by CMN and CVN.
 Hence to solve this weight estimation problem, the Expectation Maximization 
 (EM) algorithm can be used. The auxiliary function is
\begin{eqnarray}
{\bf Q}(\textbf{w},\bar{\textbf{w}}) & = & \sum^T_{t=1} \sum^M_{m=1} \gamma_m(t) \log p({\bf X}_{t}{\bf w}|\Lambda_m) \nonumber\\
        & \propto &\sum_{t,m} -\frac{1}{2}({\bf X}_{t}{\bf w}-{\bf \mu}_m){\bf \Sigma}^{-1}_m({\bf X}_{t}{\bf w}-{\bf \mu}_m)^T\nonumber\\
       & \propto & \sum^T_{t=1} -{\bf A}_t {\bf w}+{\bf B}_t
\label{eq:e2}
\end{eqnarray}
Where
\begin{eqnarray}
{\bf A}_t &=& \sum^M_{m=1}\gamma_m(t) {\bf X}_{t}\Sigma^{-1}_m {\bf X}_{t}^T  \nonumber\\
{\bf B}_t &=& \sum^M_{m=1}\gamma_m(t) {\bf X}_{t}\Sigma^{-1}_m {\bf \mu}_m
\end{eqnarray}
where $\gamma_m(t)$ is the posterior probability of the $m^{th}$ Gaussian given the observed features at frame $t$: 
\begin{equation}
\gamma_m(t)=p(m|\textbf{o}_t)=\frac{p(\textbf{o}_t|m)\omega_m}{\sum^M_{n=1}p(\textbf{o}_t|n)\omega_n}
\end{equation}
In this case, the observed features $\textbf{o}_t$ is the weighted sum ${\bf X}_{t}{\bf w}$ and $ p(\textbf{o}_t|n) $ is the 
likelihood of feature in Gaussian $n$.

The estimation of ${\bf w}$ is a least square regression problem.
The solution of (\ref{eq:e2}) can be found by:
\begin{equation}
\hat{\bf w}_t = -{\bf B}_t {\bf A}_t^{-1}
\label{eq:e3}
\end{equation}
In order to smooth the weight for each utterance, 
the mean of $\hat{\bf w}_t$ will be used as value of weight. We estimate $\hat{\bf w}_t$ of each frame then average it over the utterance.
 
The objective function in (\ref{eq:e2}) maximizes the likelihood of the weighted features
 on the clean feature’s PDF without considering the Jacobian of the weighting, although the weighting changes the feature space.
 This will cause the solution in (\ref{eq:e3}) to generally decrease the variance 
 of the weighted features due to the regression to mean effect. To address 
 this problem, we introduce 2 constraints to the weight estimation process
 in the next two sections. 

\subsection{Weight Sum Constraint}
A straightforward constraint is to map the weights to positive values summing to 1. This can be done as follows:
\begin{equation}
\tilde{ w}_c=\frac{\exp(\hat{ w}_c)}{\sum^C_{c=1}\exp(\hat{ w}_c)}
\label{eq:e4}
\end{equation}
where $\hat{ w}_c$ is the ML estimate of the weight for channel $c$, 
and $\tilde{ w}_c$ is the transformed weight. The weight transform 
in Eq. (\ref{eq:e4}) generally produces weighted features with reasonable dynamic range, however, it no longer maximize the likelihood objective function any more. 
 
\subsection{Jacobian constraint}
A more direct way to address the variance shrinking problem is 
by introducing a Jacobian term directly into the objective function as follows:
\begin{equation}
\hat{\bf w} = \argmaxB_{\bf w} \frac{1}{T}\sum^T_{t=1} \log p({\bf X}_{t}{\bf w} |\Lambda) + \frac{\beta}{2}\log|\hat{\bf C}|
\label{eq:e5}
\end{equation}
where $\hat{\bf C}=\frac{1}{T}\sum^T_{t=1}({\bf X}_{t}\textbf{w}-{\bf \mu})({\bf X}_{t}\textbf{w}-{\bf \mu})^T$ and $\mu=\frac{1}{T}\sum^T_{t=1}{\bf X}_{t}\textbf{w}$ are the sample covariance matrix and sample mean vector of the weighted features. 
The log determinant term in Eq. (\ref{eq:e5}) will create larger variance of the weighted
 features, hence preventing variance from shrinking. 

Note that when $\beta = 1$, the log determinant term $\frac{\beta}{2}\log|\hat{\bf C}|$ is the same as the 
Jacobian compensation \cite{c21,c22} used in vocal length normalization. It is also shown in \cite{c23} that 
the log determinant term appears naturally when we use a minimum Kullback Leibler (KL) divergence 
based objective function to estimate feature adaptation parameters.

There is no close form solution to the maximization problem of Eq. (\ref{eq:e5}), 
hence gradient based method can be used. Specifically, we use 
L-BFGS method \cite{c24} to find the solution of weights iteratively. 
The gradient of log determinant is presented as follow:
\begin{equation}
\frac{d\log |\hat{\bf C}|}{d w_c} = Tr ({\bf C}^{-1}({\bf D}_c+{\bf D}_c^T))
\end{equation}
where ${\bf D}_c=\frac{1}{T}\sum^T_{t=1}{\bf X}_{t}{\bf w}{\bf x}_c^T(t)$. $Tr(\bullet)$ is 
the trace of a matrix. ${\bf x}_c(t)$ is the feature of channel $c$ at frame $t$. 

\section{Experiments}
\subsection{Experimental setting}
The proposed methods are evaluated on the 3rd CHiME Speech Separation and Recognition 
Challenge (CHiME 3)\cite{c9}. The task is designed around the Wall Street Journal
 corpus and features talkers speaking in challenging noisy environments recorded using
 a 6-channel microphone array. 
The task consists of simulated and real data, each type of data has 3 sets 
(training, development and evaluation). The training data is used for acoustic model training. 
The development data is used for validating the system design during system building phase, while the evaluation data is for final evalaution. 
The real data is recorded in real noisy
 environments (on a bus, cafe, pedestrian area, and street junction) uttered by
 actual talkers. The simulated data is generated by mixing clean speech data with 
 noisy backgrounds. The corpus contains many short utterances, and the average length of utterances is about 7s. The distribution of the data sets is shown in Table~\ref{t1}.
 
\begin{table}[tb]%
\setbox0\hbox{\verb/\documentclass/}%
\caption{The number of utterances in data set of CHiME 3 \box0.}
\label{t1}
\begin{center}
\begin{tabular}{l|l|l|l}
\hline
data set & training & development & evaluation \\
\hline
simulate & 7138 & 1640 & 1320 \\
real & 1600 & 1640 & 1320 \\
\hline
\end{tabular}%
\end{center}
\end{table}

Each utterance is recorded by a 6-channel microphone array and sample synchronized. 
We name them channel 1-6. Beside 6-channel microphone array, the uttered speech 
is captured by a close talk microphone and is named by channel 0. A special characteristics 
of the real data is that channel 2 is facing the opposite direction of the speaker, hence it 
mainly captures the background noise and has a much lower SNR than other channels. However, this is not true for simulated data. 
The ASR back-end setting of our system use the same one of CHiME 3 task except for a few changes

\begin{itemize}
\item
Features: we use 40-dimension log Mel filterbank features with delta and acceleration 
features concatenated to the static features to form a 120 dimensional feature vector.
Utterancewise CMN is applied to the static features. The input of the DNN 
acoustic model is 11 frames of consecutive frames, hence the DNN has 1320 input dimensions.

\item
We use all the 6 channels to train a single acoustic model to 
produce robust acoustic model. We do not perform beamforming
 or channel weighting on training data. The motivation of our training 
 is to produce a robustly trained DNN with a large amount of training data
 with different variations, similar to data augmentation. 

\end{itemize}
In ML based channel selection experiment, a GMM is necessary for ML based selection.
 We use the channel 0's speech of training set to train a GMM with 512 mixtures, which
 is trained by 40-dimension of log Mel filterbank features. In autoencoder based 
 experiment, the autoencoder is trained by the same feature of ML based method. 
 The input and teacher signal of autoencoder use the same feature of channel 0's speech.
 For each test utterance, one channel is selected for recognition by using the ML 
 or autoencoder based methods from channels 1-6. 
 
In the channel weighting experiments, we use the GMM used 
for ML-based channel selection. The estimated weights are used to 
combine the 6 channels' features to produce a single features 
stream for speech recognition. 
The MVDR baseline systems also uses the same features and acoustic model
 as the channel weighting/selection systems. We have shown in \cite{c25} that 
 it is not good to use MVDR during training. This is because MVDR performs very 
 well on training data (mostly simulated data) and will reduce the diversity
 of the training data and hence degrade the generalization
 capability of the trained acoustic mode. 

\begin{table}[t]
\caption{The average WER of CHiME 3 development and evaluation task (\%). All results are obtained with the same acoustic model trained from all channels. Each row represents the results of a channel or processing applied during evaluation. 
``Ch 1-6'' refers to the individual channels;
``Avg'' is the average results of the 6 channels; ``MVDR (Ch 1-6) means applying MVDR on all channels, while MVDR (no Ch2) exclude channel 2;  ``Oracle CS'': channel selection with the WER information of all sentences in all channels.}
\label{t2}
\centering
\begin{tabular}{|c|c|c|c|c|c|c|c|c|c|}
\hline
\multirow{2}{*}{\textbf{Methods}} & \multicolumn{2}{|c|}{\textbf{Development Set}} & \multicolumn{2}{|c|}{\textbf{Evaluation Set}} \\
\cline{2-5}  & Simulate & Real  & Simulate & Real  \\ \hline
Ch 1                                          & 16.01    & 14.79 & 47.99    & 24.76 \\ \hline
Ch 2                                          & 14.62                         & 43.96                      & 15.85                         & 64.76 \\ \hline
Ch 3                                          & 16.66                         & 15.06                      & 19.70                         & 27.38 \\ \hline
Ch 4                                          & 11.24                         & 13.20                      & 13.14                         & 23.96 \\ \hline
Ch 5                                          & 11.32                         & 11.97                      & 14.14                         & 20.64 \\ \hline
Ch 6                                          & 20.52                         & 12.32                      & 22.33                         & 22.90 \\ \hline
Avg.(Ch 1-6)                                  & 15.06                         & 18.55                      & 22.19                         & 30.73 \\ \hline
MVDR (Ch 1-6)                                 & 6.67                          & 12.55                      & 8.09                          & 21.91 \\ \hline
MVDR (no Ch 2)                                & 7.25                          & 11.48                      & 8.75                          & 17.83 \\ \hline
Oracle CS                                     & 8.12                          & 7.73                       & 9.79                          & 14.58 \\ \hline
\end{tabular}
\end{table}

\subsection{Baseline experiments}
The word error rates (WER) of individual channels and the MVDR beamforming are shown in Table~\ref{t2}.
From the table, we observe that the WERs of channels can be significantly different. Channel 5
 results are the best on average.  Channel 2 results are the worst for real data as microphone 
 2 is used to collect background noise and not facing the speaker. However, channel 2 results on
 simulated data are comparable with other channels as microphone is facing the speaker in simulated data.

When MVDR is applied to the 6 channels, significant performance improvement can
 be obtained on the simulated data. However, MVDR performs poorly on real data and 
 its results are even worse than those of channel 5. This is mainly because the gain of the
 target signals are quite different in the 6 channels for real data. Especially, 
 the channel 2 has very low SNR. Blindly applying MVDR beamforming does not obtain 
 good results on real data. When MVDR is applied to 5 channels (excluding channel 2),
 we see improved results on real data, but degraded results on simulated data. This is
 reasonable considering that channel 2 has different characteristics in the real and 
 simulated data. 
 
Table~\ref{t2} also shows the results of the best utterance based channel selection we can obtain.
 The best channel selection is obtained by looking at the WERs of all channels for one test utterance,
 and choose the channel with the lowest WER. This result will serve as an upper bound performance for 
 the channel selection methods. From the table, there is a large room for improvement 
 for channel selection methods.

\subsection{Channel selection and weighting experiments}
The performance of our proposed methods is shown in Table~\ref{t3} for simulated data and Table~\ref{t4} for real data.
 The results of best single channel (channel 5) and the two ways of MVDR beamforming are also shown for 
 comparison. From the Tables, we have a few observations. 
 
Channel selection does not improve the ASR performance consistently 
across all test conditions when compared to channel 5. For most test 
conditions, only marginal improvements are obtained. For some test
 conditions, channel selection even degrades the performance. 
 These results could partially be due to the channel 5 
 delivering very good results in overall. 
 
Channel weighting generally performs significantly better 
than channel selection and channel 5. Among the two
 channel weighting constraints, the Jacobian constraint consistently
 outperforms the weight sum constraint in almost all test conditions. 
 In addition, for real test data, the Jacobian constrained channel
 weighting performs equally well or even better than the MVDR beamforming without channel 2.
 For example, on real development data, the Jacobian constrained weighting (Weight-Jacobian)
 achieves an average WER of 9.97\%, while MVDR without channel 2 only produces an average WER of 11.48\%.
 These results show the advantage of channel weighting over traditional beamforming.

\begin{table}[tb]%
\caption{The WER of proposed methods evaluated on CHiME 3 \textbf{Simulate data set} (\%). Method means: Ch 5: result of channel 5; MVDR: MVDR beamforming with all channel and all channel without channel 2; Select-ML and Select-AE: Channel selection with Maximum likelihood and autoencoder; Weight-Sum and Jacobian: Channel weight with Sum to 1 and Jacobian constraint}
\label{t3}
\begin{center}
\begin{tabular}{|l|l|l|l|l|l|}
\hline
\multirow{2}{*}{Method} & \multicolumn{5}{l|}{Test Environments} \\ \cline{2-6} 
                        & BUS    & CAFE   & PED  & STRT  & Avg.  \\ \hline
\multicolumn{6}{|c|}{Development Set}                            \\ \hline
Ch 5                    & 11.52  & 14.60  & 9.13 & 10.01 & 11.32 \\ \hline
MVDR                    & 5.94   & 8.32   & 5.81 & 6.59  & 6.67  \\ \hline
MVDR(no ch2)            & 6.74   & 9.07   & 6.36 & 6.83  & 7.25  \\ \hline
Select-ML               & 9.49   & 14.23  & 9.08 & 9.21  & 10.50 \\ \hline
Select-AE               & 10.51  & 14.79  & 9.22 & 10.52 & 11.26 \\ \hline
Weight-Sum              & 16.83  & 10.71  & 7.05 & 11.33 & 11.48 \\ \hline
Weight-Jacobian         & 9.19   & 13.91  & 8.95 & 9.50  & 10.39 \\ \hline
\multicolumn{6}{|c|}{Evaluation set}                                           \\ \hline
Ch 5                    & 11.90  & 16.55  & 14.33 & 13.78 & 14.14 \\ \hline
MVDR                    & 6.78   & 9.45   & 7.68 & 8.44  & 8.09  \\ \hline
MVDR(no ch2)            & 7.34   & 9.99   & 8.52 & 9.13  & 8.75  \\ \hline
Select-ML               & 10.06  & 15.81  & 14.42 & 12.49  & 13.20 \\ \hline
Select-AE               & 10.47  & 15.83  & 13.54 & 11.95 & 12.95 \\ \hline
Weight-Sum              & 10.39  & 15.78  & 12.42 & 13.06 & 12.91 \\ \hline
Weight-Jacobian         & 10.44  & 14.38  & 11.24 & 12.98  & 12.26 \\ \hline
\end{tabular}
\end{center}
\end{table}

\begin{table}[tb]%
\caption{The WER of proposed methods evaluated on CHiME~3 \textbf{Real data set} (\%). }
\label{t4}
\begin{center}
\begin{tabular}{|l|l|l|l|l|l|}
\hline
\multirow{2}{*}{Method} & \multicolumn{5}{l|}{Test Environments} \\ \cline{2-6} 
                        & BUS    & CAFE   & PED  & STRT  & Avg.  \\ \hline
\multicolumn{6}{|c|}{Development Set}                            \\ \hline
Ch 5                    & 18.14  & 10.72  & 7.73 & 11.28 & 11.97 \\ \hline
MVDR                    & 15.39   & 11.36  & 10.43 & 13.01  & 12.55  \\ \hline
MVDR(no ch2)            & 13.94   & 10.16  & 9.68 & 12.12  & 11.48  \\ \hline
Select-ML               & 15.26   & 11.71  & 7.71 & 11.25  & 11.48 \\ \hline
Select-AE               & 16.30  & 10.81  & 7.49 & 11.93 & 11.63 \\ \hline
Weight-Sum              & 11.18  & 14.23  & 8.63 & 9.48 & 10.88 \\ \hline
Weight-Jacobian         & 13.66  & 10.38  & 6.53 & 9.29 & 9.97 \\ \hline
\multicolumn{6}{|c|}{Evaluation set}                             \\ \hline
Ch 5                    & 29.56  & 22.43  & 16.95 & 13.63 & 20.64 \\ \hline
MVDR                    & 28.10   & 24.45   & 19.39 & 15.71  & 21.91  \\ \hline
MVDR(no ch2)            & 22.40   & 19.26   & 15.19 & 14.48  & 17.83  \\ \hline
Select-ML               & 29.63  & 22.23  & 17.58 & 14.51  & 20.99 \\ \hline
Select-AE               & 28.47  & 22.15  & 16.61 & 14.15 & 20.35 \\ \hline
Weight-Sum              & 26.55  & 19.56  & 15.58 & 14.05 & 18.94 \\ \hline
Weight-Jacobian         & 25.80  & 18.68  & 14.09 & 12.40 & 17.74 \\ \hline
\end{tabular}
\end{center}
\end{table}

\begin{figure}[tb]
\begin{center}
\includegraphics[scale=0.3]{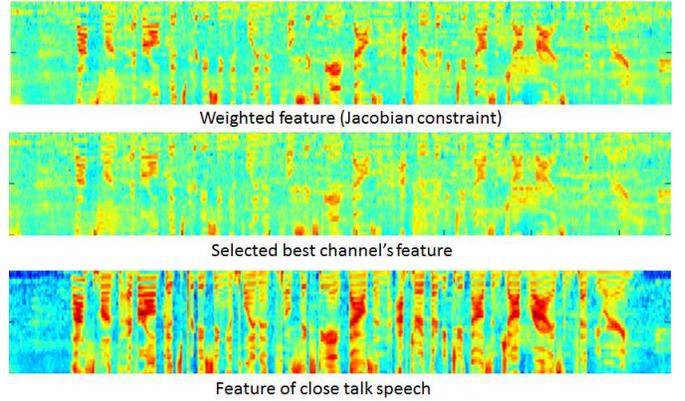}
\end{center}
\caption{The log filter bank graph of weighted, selected and close talk feature for one utterance.
The horizontal axis is for time, and vertical axis is for different frequency band.}
\label{fig:2}
\end{figure}

We also show one example of selected/weighted log Mel filterbanks 
in Fig. \ref{fig:2}. From the figure, we can see that the weighted features
 (by Jacobian constrained channel weighting) looks more similar to the 
 close-talk feature than the best channel 
 (For this utterance, channel 5 is selected as best channel). This is because there is higher flexibility
 in channel weighting than that of channel selection. 

\section{Conclusion and future work}

In this paper, we studied two channel selection and two channel weighting methods 
for robust ASR with multiple microphone channels. Experimental results on the 
CHiME 3 task show promising results of the channel weighting. In the future, 
we will extend the channel weighting method in several ways. For example, 
we can estimate separated weight for each filterbank rather than using a 
single weight for all filterbanks, as some channel may have good 
quality in some filterbank, although its overall quality is not
 good. It is also possible to make the weights slowly varying with time.

\section*{Acknowledgments}

This work was supported by JSPS KAKENHI Grant Number
15K16020 and a research grant from the Research Foundation
for the Electrotechnology of Chubu (REFEC).


\profile[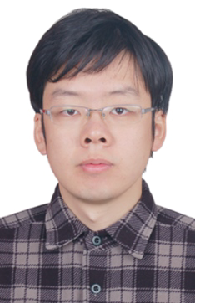]{Zhaofeng Zhang}{%
received the B.S. degrees in Northwest University in 2010, China. M.S. degrees in Department 
of System Engineering, Shizuoka University in 2013, Japan. During 2014-present, 
he is staying at Nagaoka University of Technology in Niigata, Japan, to study noisy robust 
speech/speaker recognition technical for his PhD course. 
His research interests include robust speech/speaker recognition and speaker verification.
}
\label{profile}

\profile[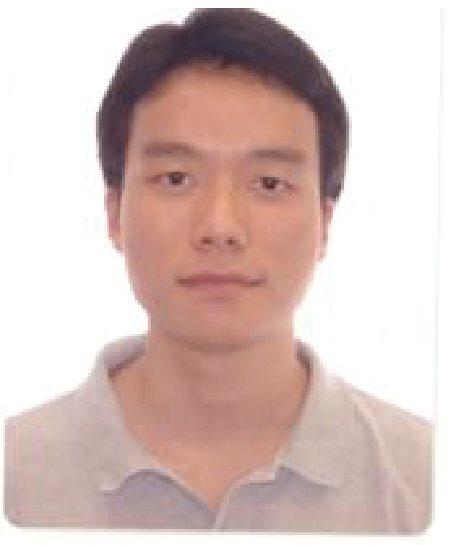]{Xiong Xiao}{received his B.Eng and Ph.D degrees in computer engineering from Nanyang Technological University (NTU) in 2004 and 2010, respectively. He joined Temasek laboratories @ NTU in 2009 and is now a senior research scientist. 
His research interests include robust speech processing, spoken document retrieval, and signal processing. 

}

\profile[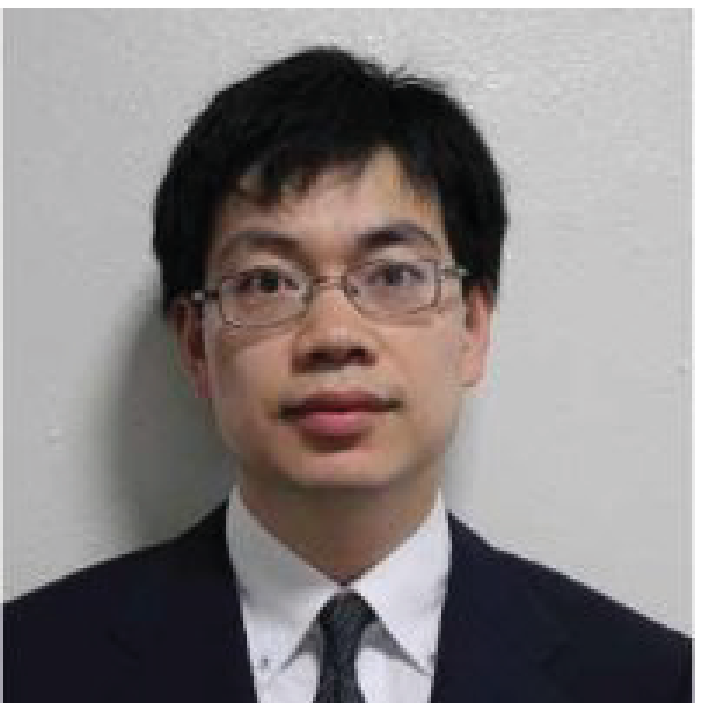]{Longbiao Wang}{%
received his B.E. degree from Fuzhou University, China, in 2000 
and an M.E. and Dr.　Eng. degree from Toyohashi University of Technology, 
Japan, in 2005 and 2008, respectively. From July 2000 to August 2002, 
he worked at the China Construction Bank. He was an Assistant Professor in 
the faculty of Engineering at Shizuoka University, Japan, from April 2008 to September 2012.
 Since October 2012, he has been an Associate Professor at Nagaoka University of Technology, Japan.
 His research interests include robust speech recognition, speaker recognition and sound source 
 localization. He is a member of the IEEE, the Institute of Electronics, Information and 
 Communication Engineers (IEICE) and the Acoustical Society of Japan (ASJ).
}

\profile[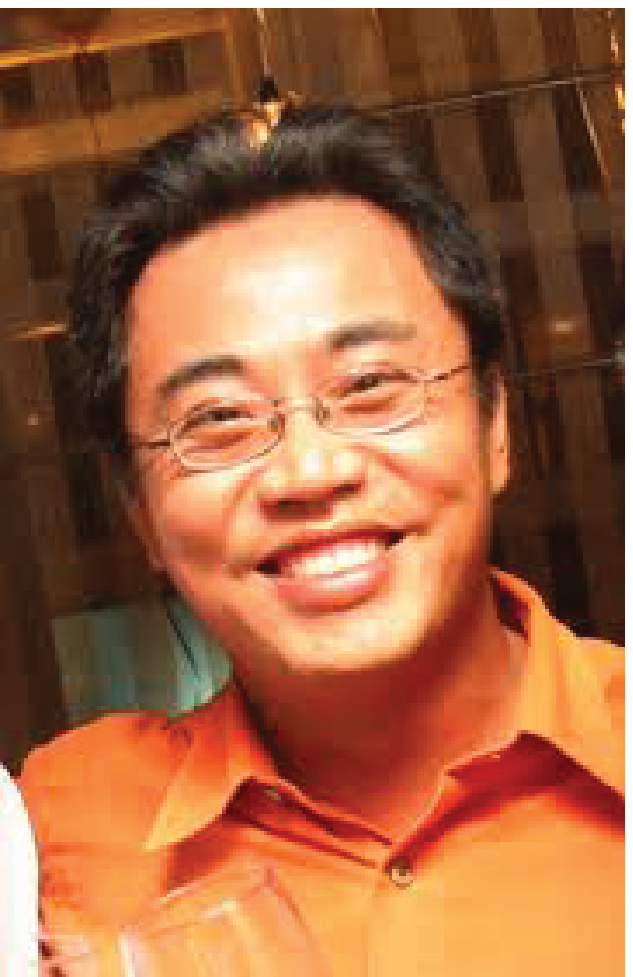]{Eng Siong Chng}{
received the B.Eng. (honors) degree in electrical and electronics engineering and the Ph.D. degree from the University of Edinburgh, Edinburgh, U.K., in 1991 and 1996, respectively. He is currently an Associate Professor in the School of Computer Engineering, Nanyang Technological University (NTU), Singapore. Prior to joining NTU in 2003, he was with the Institute of Physics and Chemical Research, Riken, as a Postdoctoral Researcher working in the area of signal processing and classification (1996), the Institute of System Science (ISS, currently known as I$^2$R) as a member of research staff to transfer the Apple-ISS speech and handwriting technologies to ISS (1996-1999), Lernout and Hauspie (now part of Nuance) as a Senior Researcher in speech recognition (1999-2000), and Knowles Electronics as a Manager for the Intellisonic microphone array research (2001-2002). His research interests are in pattern recognition, signal, speech, and video processing. His research interests are in pattern recognition, signal, speech and video processing. He has published over 50 papers in international journals and conferences.  He is currently leading the speech and language technology program (http://www3.ntu.edu.sg/home/aseschng/SpeechTechWeb/default.htm) in Emerging Research Lab at the School of Computer Engineering, NTU. 
}

\profile[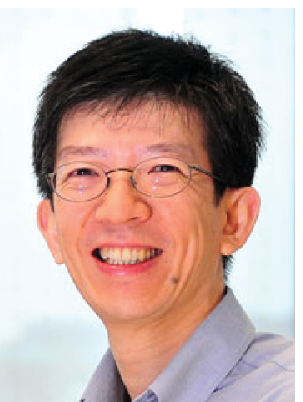]{Haizhou Li}{
received the B.Sc., M.Sc., and Ph.D degree in electrical and electronic engineering from South China University of Technology, Guangzhou, China in 1984, 1987, and 1990 respectively. Dr Li is currently the Principal Scientist, Department Head of Human Language Technology in the Institute for Infocomm Research (I$^2$R), Singapore. He is also an adjunct Professor at the Nanyang Technological University, National University of Singapore and the University of New South Wales, Australia. His research interests include automatic speech recognition, speaker and language recognition, natural language processing, and computational intelligence.

Prior to joining I$^2$R, he taught in the University of Hong Kong (1988-1990) and South China University of Technology (1990-1994). He was a Visiting Professor at CRIN in France (1994-1995), a Research Manager at the Apple-ISS Research Centre (1996-1998), a Research Director in Lernout \& Hauspie Asia Pacific (1999-2001), and the Vice President in InfoTalk Corp. Ltd. (2001-2003). 

Dr Li is currently the Editor-in-Chief of IEEE/ACM Transactions on Audio, Speech and Language Processing (2015-2017), a Member of the Editorial Board of Computer Speech and Language (2012-2016), the President of the International Speech Communication Association (2015-2017), and the President of Asia Pacific Signal and Information Processing Association (2015-2016).  He was an elected Member of IEEE Speech and Language Processing Technical Committee (2013-2015), the General Chair of ACL 2012, and INTERSPEECH 2014. 

Dr Li is a Fellow of the IEEE. He was a recipient of the National Infocomm Award 2002 and the President's Technology Award 2013 in Singapore. He was named one the two Nokia Visiting Professors in 2009 by the Nokia Foundation. }

\end{document}